\documentclass[draft,tightenlines,nofootinbib,preprint,aps,eqsecnum,amsmath,amssymb]{revtex4}

\begin{document}

\newcommand{\sgn}{\mbox{\boldmath $\epsilon$}}

\newcommand{\eo}{{}^4{\buildrel \circ \over E}}

\newcommand{\beq}{\begin{equation}}
\newcommand{\eeq}{\end{equation}}
\newcommand{\bea}{\begin{eqnarray}}
\newcommand{\eea}{\end{eqnarray}}
\newcommand{\cir}{{\buildrel \circ \over =}}

\newcommand{\on}{\stackrel{\circ}{=}}
\newcommand{\byd}{\stackrel{def}{=}}
\baselineskip 20pt

\title{Killing Symmetries as Hamiltonian Constraints.}

\author{Luca Lusanna}

\affiliation
{Sezione INFN di Firenze,
Via G. Sansone 1\\
50019 Sesto Fiorentino (FI), Italy\\
E-mail: lusanna@fi.infn.it}

\begin{abstract}

The existence of a Killing symmetry in a gauge theory is equivalent to the addition of extra
Hamiltonian constraints in its phase space formulation, which imply restrictions both on the Dirac
observables (the gauge invariant physical degrees of freedom) and on
the gauge freedom.

When there is a time-like Killing vector field only pure gauge electromagnetic
fields survive in Maxwell theory in Minkowski space-time , while in ADM canonical gravity
in asymptotically Minkowskian space-times only inertial effects without gravitational waves
survive.

\today

\end{abstract}

\maketitle

\newpage

\section{Introduction}

Killing symmetries are of basic importance in the search
of exact solutions of Einstein's equations. If the metric tensor
satisfies $L_X\, {}^4g_{\mu\nu} = 0$, the vector field $X =
\xi^{\mu}\, \partial_{\mu}$ is a Killing vector field (satisfying
the  8 Killing equations $\nabla_{\mu}\, \xi_{\nu} + \nabla_{\nu}\,
\xi_{\mu} = \partial_{\mu}\, \xi_{\nu} +
\partial_{\nu}\, \xi_{\mu} - 2\, {}^4\Gamma^{\alpha}_{\mu\nu}\,
\xi_{\alpha} = 0$) and the space-time has a Killing symmetry and
privileged 4-coordinates adapted to it leading to a simplification
of Einstein equations.

\bigskip

It is known from the work in Refs.\cite{1,2,3,4,5}  that for Einstein and
Eistein-Maxwell theories \footnote{Einstein-Yang-Mills is even more
complicated due to the gauge symmetries, i.e. the Gribov ambiguity,
of the Yang-Mills connection \cite{6,7}.} the presence of Killing
symmetries, when the tensors over space-time belong to ordinary
Sobolev spaces, introduces singularities in the space of 4-metrics.
This space is not a manifold but has a {\it cone over cone}
structure of singularities: there is a cone of 4-metrics with a
Killing symmetry, from each point of this cone emerges a cone of
4-metrics with two Killing symmetries and so on.
Choquet-Bruhat \cite{8,9}  has shown that  if the tensors belong to certain weighted Sobolev
spaces \footnote{They have been introduced to validate tensor
decompositions like the Hodge one at spatial infinity in
asymptotically flat space-times. See Ref.\cite{10}.} then these
singularities disappear, because the implied boundary conditions on
the tensors after a 3+1 splitting of the space-time exclude the existence of Killing vectors
\footnote{Analogously Moncrief \cite{6}  has shown that gauge
symmetries and Gribov ambiguity are absent in certain weighted
Sobolev spaces.}. Therefore if all the fields  belong to suitable {\it
weighted Sobolev spaces} then:\hfill\break
 i) the admissible space-like hyper-surfaces (the 3-spaces of a 3+1 splitting) are Riemannian 3-manifolds
without asymptotically vanishing Killing vectors \cite{8,9}; as a
consequence of the assumed boundary conditions no Killing vectors
can be present except the asymptotic Killing symmetries (the 10 ADM
Poincare' generators \cite{11}  of the asymptotic Poincare' group)
in the case of asymptotically Minkowskian space-times;\hfill\break
 ii) the inclusion of particle physics leads to a formulation without
Gribov ambiguity \cite{6,7}.
\bigskip

As a consequence, in this class of space-times the spaces of
4-metrics and of 3-metrics (after a 3+1 splitting) should be smooth
manifolds without singularities.
If a space-time of this class without non-asymptotic Killing symmetries and with
the fields belonging to suitable weighted Sobolev spaces is globally hyperbolic,
topologically trivial, asymptotically Minkowskian and without super-translations \cite{12}
then there is a well established Hamiltonian description (using Dirac's theory of constraints \cite{13,14})
of both metric and tetrad gravity \cite{15,16,17,18,19,20,21}
(see Refs.\cite{22}  for  reviews). This is due to the fact that
in these space-times one can make a consistent 3+1 splitting with instantaneous 3-spaces (i.e. a clock
synchronization convention) centered on a time-like observer used as origin of (world scalar)
radar 4-coordinates \cite{23,24}: in this way the notion of non-inertial frames defined in Minkowski space-time in Refs. \cite{25}
can be extended to this class of curved space-times.
The absence of super-translations implies that the SPI group \cite{26} of asymptotic
Killing symmetries is reduced to the asymptotic ADM Poincar\'e group \cite{15,16}.

\bigskip

To our knowledge there is no Hamiltonian treatment of Killing symmetries. In this paper
we try to see what happens if we relax the hypothesis of weighted Sobolev
spaces and we add by hand the ten Killing equations corresponding to
a {\it given} Killing vector field $X = \xi^{\mu}\, \partial_{\mu}$
or $X = \xi^{\tau}\, \partial_{\tau} + \xi^r\, \partial_r$ (in radar
4-coordinates adapted to the 3+1 splitting). The ten Killing equations are
restrictions on the 4-metric forcing it to belong to the singularity cone
of metrics with one Killing vector.

\medskip

To implement this program we have to rewrite the 10 Killing
equations for the given $X$ as Hamiltonian Dirac constraints to be
added by hand to the set  of (8 or 14) first class constraints of
metric or tetrad gravity. These extra constraints restrict the
constraint manifold in phase space to contain only metrics with the
given Killing vector field $X$ and subsequently will reduce the
space of solutions of Hamilton and Einstein equations. A consistent
theory will emerge if the extra constraints are compatible with the
existing ones. To check this point we have to ask for the
time-preservation of the extra constraints: their Poisson bracket
with the Dirac Hamiltonian \footnote{The extra constraints must not be added
to the Dirac Hamiltonian with Dirac multipliers as we do with
primary constraints, because they are interpreted as restrictions on
the field configurations added by hand. Only if one would find a
singular Lagrangian invariant under local gauge transformations generated
by a Killing symmetry, one could use the standard Dirac algorithm with all
the Killing constraints already present.} must vanish.

\medskip

Since the ten Killing constraints corresponding to a Killing vector
must be preserved in time, other 10 constraints emerge (they too
should be preserved in time, but this should not add new
constraints). Therefore each Killing vector gives rise to twenty
extra constraints identifying a well defined sub-manifold of the
constraint manifold.

\medskip

If there are no pathologies, the net effect of the Killing symmetry will be
to impose a symmetry pattern on the tidal variables (the physical degrees of
freedom of the gravitational field becoming the two polarizations of the gravitational
waves in the linearized theory) and on the inertial ones (describing the gauge freedom
in the choice of the coordinates). In particular we
will show that a time-like Killing vector eliminates completely the tidal degrees of freedom, so
that the resulting space-time contains only stationary inertial effects and maybe singularities
like black holes. Extra Killing symmetries should only restrict these residual inertial
effects.

\bigskip

Before treating gravity, we will show what is the effect of symmetry
on the electro-magnetic field in Minkowski space-time. This will be done by adding the requirement that the
connection $A_{\mu}$ satisfies the symmetry requirement $L_X\,
A_{\mu} = 0$ for a given vector field $X = \xi^{\mu}\,
\partial_{\mu}$. Now four constraints are associated to each such
vector: their time preservation generates other four constraints, so
that each  symmetry has eight associated constraints.
We will see that in the case of a time-like symmetry vector field
only  {\it pure gauge} electro-magnetic potentials survive: there
are no radiative degrees of freedom (eventually a singularity in the
case of magnetic monopoles could be present). With other types of
symmetry vector fields the radiative degrees of freedom (and also the gauge
freedom) are forced to obey the symmetry.

\bigskip

In Section II we describe the 3+1 splittings of the quoted space-times and
the notion of radar 4-coordinates, which allows to give a description of gravity
in terms of world-scalar quantities  in non-inertial frames.

In Section III, after remembering the formulation \cite{25} of the Hamiltonian version of electro-magnetisnm
in non-inertial frames and in the inertial rest frame in Minkowski space-time, we study the Hamiltonian version
of  symmetry conditions (special relativistic Killing symmetries) on the electro-magnetic field. The case of a
time-like symmetry is treated explicitly.

In Section IV, after remembering the formulation \cite{20} of Hamiltonian tetrad gravity in the
non-inertial frames of the quoted space-times, we give
the Hamiltonian formulation of Killing symmetries in the York canonical basis of
tetrad gravity and we study the implications of the presence of a time-like Killing vector.

In  Appendix A we give the Hamiltonian expression of the extrinsic curvature and of the Christoffel symbols.

\section{3+1 Splitting and Radar 4-Coordinates}

Assume that the world-line $x^{\mu}(\tau)$ of an arbitrary time-like
observer carrying a standard atomic clock is given either in Minkowski space-time
or in the quoted class of Einstein space-times: $\tau$ is an arbitrary
monotonically increasing function of the proper time of this clock. Then one
gives an admissible 3+1 splitting of the asymptotically flat space-time, namely a nice
foliation with space-like instantaneous 3-spaces $\Sigma_{\tau}$. It is the
mathematical idealization of a protocol for clock synchronization: all the
clocks in the points of $\Sigma_{\tau}$ sign the same time of the atomic
clock of the observer. The observer and the
foliation define a global non-inertial reference frame after a choice of
4-coordinates. On each 3-space $\Sigma_{\tau}$  one chooses curvilinear
3-coordinates $\sigma^r$ having the observer as origin.
The quantities $\sigma^A = (\tau; \sigma^r)$ are the either Lorentz- or world-scalar and observer-dependent
\textit{radar 4-coordinates}, first introduced by Bondi \cite{23,24}. \medskip

If $x^{\mu} \mapsto \sigma^A(x)$ is the coordinate transformation from
world 4-coordinates $x^{\mu}$ having the observer as origin to radar 4-coordinates, its inverse $\sigma^A
\mapsto x^{\mu} = z^{\mu}(\tau ,\sigma^r)$ defines the \textit{embedding}
functions $z^{\mu}(\tau ,\sigma^r)$ describing the 3-spaces $\Sigma_{\tau}$
as embedded 3-manifolds into the asymptotically flat space-time.
Let $z^{\mu}_A(\tau, \sigma^u) = \partial\, z^{\mu}(\tau, \sigma^u) / \partial\, \sigma^A$ denote the gradients of the
embedding functions with respect to the radar 4-coordinates.
The space-like 4-vectors $z^{\mu}_r(\tau ,\sigma^u)$ are tangent to $\Sigma_{\tau}$, so that the unit
time-like normal $l^{\mu}(\tau ,\sigma^u)$ is proportional to $\epsilon^{\mu}{}_{%
\alpha \beta\gamma}\, [z^{\alpha}_1\, z^{\beta}_2\, z^{\gamma}_3](\tau
,\sigma^u)$ ($\epsilon_{\mu\alpha\beta\gamma}$ is the Levi-Civita tensor).
Instead $z^{\mu}_{\tau}(\tau, \sigma^u)$ is a time-like 4-vector skew with respect to the 3-spaces leaves of the foliation.
In special relativity (SR), see Refs. \cite{25},
one has $z^{\mu}_{\tau}(\tau ,\sigma^r) = [N\, l^{\mu} + N^r\,
z^{\mu}_r](\tau ,\sigma^r)$ with $N(\tau ,\sigma^r) = \epsilon\,
[z^{\mu}_{\tau}\, l_{\mu}](\tau ,\sigma^r) = 1 + n(\tau, \sigma^r) > 0$ and $
N_r(\tau ,\sigma^r) = - \epsilon\, [z^{\mu}_{\tau}\, \eta_{\mu\nu}\, z_r^{\mu}](\tau ,\sigma^r)$ being the
lapse and shift functions respectively of the {\it global non-inertial frame} of Minkowski space-time so defined.\medskip

In SR the classical fields, for instance the Klein-Gordon field $\tilde \phi(x^{\mu})$, have
to be replaced with fields knowing the 3+1 splitting: $\phi(\tau, \sigma^r) = \tilde \phi(z^{\mu}(\tau, \sigma^r))$.
With parametrized Minkowski theories \cite{25,27}, one can give a Lagrangian formulation of classical fields
in non-inertial frames with a Lagrangian depending also on the embedding variables $z^{\mu}(\tau, \sigma^r)$.
The resulting action is invariant under frame preserving diffeomorphisms. As a consequence the embedding
variables are gauge variables and the transition from a non-inertial frame to another (either non-inertial or
inertial) one is a gauge transformation. Inertial frames are a special case of this description. For every isolated
system with a conserved time-like total 4-momentum $P^{\mu}$ one can define the {\it inertial rest frame} as the one
in which the Euclidean 3-spaces are orthogonal to $P^{\mu}$.

\medskip

In general relativity (GR) the dynamical fields are the components ${}^4g_{\mu\nu}(x)$ of
the 4-metric and not the  embeddings $x^{\mu} = z^{\mu}(\tau,
\sigma^r)$ defining the admissible 3+1 splittings of space-time. Now the gradients
$z^{\mu}_A(\tau, \sigma^r)$ of the embeddings give the transition
coefficients from radar to world 4-coordinates, so that the
components ${}^4g_{AB}(\tau, \sigma^r) = z^{\mu}_A(\tau, \sigma^r)\,
z^{\nu}_B(\tau, \sigma^r)\, {}^4g_{\mu\nu}(z(\tau, \sigma^r))$ of
the 4-metric will be the dynamical fields in the ADM action \cite{11}.
\medskip

Let us remark that {\it the ten quantities ${}^4g_{AB}(\tau, \sigma^r)$ are
4-scalars of the space-time due to the use of the world-scalar radar 4-coordinates}. In each
3-space $\Sigma_{\tau}$ considered as a 3-manifold with 3-coordinates $\sigma^r$ (and
not as a 3-sub-manifold of the space-time) ${}^4g_{\tau r}(\tau, \sigma^u)$ is a 3-vector and ${}^4g_{rs}(\tau, \sigma^u)$ is a 3-tensor.
 Therefore {\it all the components of  "radar tensors", i.e. tensors expressed in radar 4-coordinates, are
4-scalars of the space-time} \cite{21}.
\medskip

In the considered class of Einstein space-times the ten {\it strong} asymptotic ADM
Poincar\'e generators $P^A_{ADM}$, $J^{AB}_{ADM}$ (they are fluxes
through a 2-surface at spatial infinity) are well defined
functionals of the 4-metric fixed by the boundary conditions at
spatial infinity and of matter (when present). These ten strong generators can be
expressed \cite{15,16}  in terms of the weak asymptotic ADM Poincar\'e generators (integrals on
the 3-space of suitable densities) plus first class constraints.
The absence of super-translations implies that the ADM 4-momentum is asymptotically orthogonal to the
instantaneous 3-spaces (they tend to a Euclidean 3-space at spatial infinity). As a consequence each
3-space of the global non-inertial frame is a {\it non-inertial rest frame} of the 3-universe. At spatial
infinity there are asymptotic inertial observers carrying a flat tetrad whose spatial axes are identified by the fixed stars of star catalogues.

\bigskip

The 4-metric ${}^4g_{AB}$ has signature $\sgn\, (+---)$ with $\sgn =
\pm$ (the particle physics, $\sgn = +$, and general relativity,
$\sgn = -$, conventions). Flat indices $(\alpha )$, $\alpha = o, a$,
are raised and lowered by the flat Minkowski metric
${}^4\eta_{(\alpha )(\beta )} = \sgn\, (+---)$. We define
${}^4\eta_{(a)(b)} = - \sgn\, \delta_{(a)(b)}$ with a
positive-definite Euclidean 3-metric. From now on we shall denote
the curvilinear 3-coordinates $\sigma^r$ with the notation $\vec
\sigma$ for the sake of simplicity. Usually the convention of sum
over repeated indices is used, except when there are too many
summations. The symbol $\approx$ means Dirac weak equality, while the symbol
$\cir$ means evaluated by using the equations of motion.

\section{Electro-Magnetism}

In SR the non-dynamical Minkowski space-time
admits 10 Killing vectors, the generators of the algebra of the
kinematical Poincare' group connecting inertial frames. Let us see
what happens to the physical degrees of freedom, the Dirac
observables (DO), of the electro-magnetic field in absence of matter,
when the electro-magnetic potential is assumed to be left invariant,
i.e. ${\cal L}_X\, A = 0$, by anyone of the 10 Poincare' generators.
If matter is present, also the matter fields must be assumed to be
invariant under the action of the Killing vector field $X$. In other
words, let us look at what kind of restrictions on the phase space of
the electro-magnetic field are implied by anyone of
these Killing symmetries. This will be done in the inertial rest frame of the electro-magnetic field
after a review of its Hamiltonian formulation..

\subsection{Canonical Basis for Electro-Magnetism in the Inertial Rest Frame of Minkowski Space-Time}

Let us consider Dirac's Hamiltonian formulation \cite{28}  of the
electro-magnetic field, reformulated \cite{25}  in the rest-frame
instant form of dynamics on the Wigner 3-space $\Sigma_{\tau}$, i.e. on
the space-like hyper-plane orthogonal to the
conserved 4-momentum of the isolated system formed by the
electro-magnetic field ($P^{\mu}$ is the conserved 4-momentum of the
field configuration). This is an inertial frame centered on the
covariant Fokker-Pryce center of inertia and uses radar
4-coordinates $(\tau ,\vec \sigma )$. In this formulation there is no breaking of covariance, since all
the quantities on a Wigner 3-space are either Lorentz scalars or
Wigner spin 1 3-vectors. The Wigner 3-space $\Sigma_{\tau}$ at
time $\tau $ is the intrinsic rest frame of the isolated system at
time $\tau $. With respect to an arbitrary inertial frame the Wigner
hyper-planes are described by the following embedding

\begin{equation}
z^{\mu }(\tau ,\vec{\sigma})=x_{s}^{\mu }(\tau )+\epsilon _{r}^{\mu
}(u(P))\sigma ^{r},
  \label{3.1}
\end{equation}

\noindent with $x_{s}^{\mu }(\tau ) = x^{\mu}(0) + u^{\mu}(P)\,
\tau$ being the world-line of the Fokker-Pryce inertial observer.
The space-like 4-vectors $\epsilon _{r}^{\mu }(u(P))$ together with
the time-like one $\epsilon _{o}^{\mu }(u(P))$ are the columns of
the standard Wigner boost for time-like Poincare' orbits that sends
the time-like four-vector $P^{\mu }$ to its rest-frame form
$\overset{\circ }{P} {}^{\mu }=\sqrt{P^{2}}(1;\vec{0})$:
$\epsilon _{o}^{\mu }(u(P))=u^{\mu }(P)=P^{\mu }/\sqrt{P^{2}}$,
$\epsilon _{r}^{\mu }(u(P))= \Big(-u_{r}(P);\delta
_{r}^{i}-{\frac{{u^{i}(P)\,u_{r}(P)}}{{ 1+u^{o}(P)}}}\Big)$.

\bigskip

The configuration variable is  the Lorentz-scalar electro-magnetic
potential $A_A(\tau ,\vec \sigma ) = z^{\mu}_A(\tau, \vec \sigma)\, {\tilde
A}_{\mu}(z^{\beta}(\tau ,\vec \sigma ))$, whose associated field
strength is $F_{AB}(\tau ,\vec \sigma ) = \partial_A\, A_B(\tau ,\vec \sigma ) -
\partial_B\, A_A(\tau ,\vec \sigma ) = z^{\mu}_A(\tau ,\vec \sigma
)\, z^{\nu}_B(\tau ,\vec \sigma )\, {\tilde F}_{\mu\nu}(z^{\beta}(\tau ,\vec
\sigma ))$. The conjugate momentum variables are a scalar
$\pi^{\tau}(\tau ,\vec \sigma ) \approx 0$ and a Wigner 3-vector
$\pi^r(\tau ,\vec \sigma ) = E^r(\tau ,\vec \sigma )$.
$E^r(\tau ,\vec \sigma )$ and $B^r(\tau ,\vec \sigma )$ are the components of
the electric and magnetic fields in the inertial rest frame.
\medskip

The gauge degrees of freedom ($A_{\tau}$, $\eta$) have been
separated from the transverse DO's ($A_{\perp r}$, $
\pi^r_{\perp} = E^r_{\perp}$) ($\vec \partial \cdot {\vec A}_{\perp} =
\vec \partial \cdot {\vec \pi}_{\perp} = 0$) by means of a Shanmughadhasan
canonical transformation \cite{29,30}  adapted to the two scalar first
class constraints ($\pi^{\tau}(\tau ,\vec \sigma ) \approx 0$ and the Gauss law $\Gamma(\tau ,\vec \sigma )
= \vec \partial \cdot {\vec \pi}(\tau ,\vec \sigma ) \approx 0$) generators of the
Hamiltonian electro-magnetic gauge transformations ($\triangle = -
{\vec \partial}^2_{\sigma}$, $\Box = \partial^2_{\tau} + \triangle$)

\begin{eqnarray*}
 &&\begin{minipage}[t]{1cm}
\begin{tabular}{|l|} \hline
$A_A$ \\  \hline
 $\pi^A$ \\ \hline
\end{tabular}
\end{minipage} \ {\longrightarrow \hspace{.2cm}} \
\begin{minipage}[t]{2 cm}
\begin{tabular}{|l|l|l|} \hline
$A_{\tau}$ & $\eta$   & $A_{\perp\, r}$   \\ \hline
$\pi^{\tau}\approx 0$& $\Gamma \approx 0$ &$\pi^r_{\perp}$ \\
\hline
\end{tabular}
\end{minipage}
 \end{eqnarray*}

 \bea
 A_r(\tau ,\vec \sigma )&=&
\partial_r\, \eta (\tau ,\vec \sigma
)+A^r_{\perp}(\tau ,\vec \sigma ),\qquad
 \pi^r(\tau ,\vec \sigma ) = \pi^r_{\perp}(\tau
,\vec \sigma )+{1\over {\triangle_{\sigma}} }{ {\partial}\over
{\partial \sigma^r}} \, \Gamma (\tau ,\vec \sigma ), \nonumber \\
 &&{}\nonumber \\
  \eta(\tau ,\vec \sigma )&=&-{1\over {\triangle_{\sigma}} }{ {\partial} \over
 {\partial \vec \sigma} }\cdot \vec A(\tau ,\vec \sigma ),\nonumber \\
 &&{}\nonumber \\
 A^r_{\perp}(\tau ,\vec \sigma ) &=& (\delta^{rs}+{{\partial^r_{\sigma}\partial^s_{\sigma}}
 \over {\triangle_{\sigma} }})\, A_s(\tau ,\vec \sigma ),\qquad
  \pi^r_{\perp}(\tau ,\vec \sigma ) = (\delta^{rs}+{{\partial^r_{\sigma}\partial^s_{\sigma}}
 \over {\triangle_{\sigma} }})\, \pi_s(\tau ,\vec \sigma ),\nonumber \\
 &&{}\nonumber \\
  &&\lbrace A_{\tau}(\tau ,\vec \sigma ), \pi^{\tau}
(\tau ,{\vec \sigma}^{'} ) \rbrace = -
  \lbrace \eta (\tau ,\vec \sigma ),\Gamma
(\tau ,{\vec \sigma}^{'} ) \rbrace =  \delta^3(\vec
\sigma -{\vec \sigma}^{'}),\nonumber \\
 &&{}\nonumber \\
 &&\lbrace A^r_{\perp}(\tau ,\vec
\sigma ),\pi^s_{\perp}(\tau ,{\vec \sigma} ^{'})\rbrace =-
(\delta^{rs}+{{\partial^r_{\sigma}\partial^s_{\sigma}} \over
{\triangle_{\sigma} }})\delta^3(\vec \sigma -{\vec \sigma}^{'}).
 \label{3.2}
 \eea

The Dirac Hamiltonian is ($\lambda_{\tau}(\tau ,\vec \sigma )$ is
the arbitrary Dirac multiplier associated to the primary constraint
$\pi^{\tau}(\tau ,\vec \sigma ) \approx 0$)

 \bea
 H_D &=& H_c + \int d^3\sigma\, [\lambda_{\tau}\, \pi^{\tau}-
 A_{\tau}\, \Gamma ](\tau ,\vec \sigma ),\qquad
 H_c = {1\over 2}\, \int d^3\sigma\, [{\vec \pi}^2_{\perp}
 + {\vec B}^2](\tau ,\vec \sigma ),\nonumber \\
 &&{}\nonumber \\
 \Downarrow &&kinematical\, Hamilton \, equations\nonumber \\
 &&{}\nonumber \\
 \partial_{\tau}\, A_{\tau}(\tau ,\vec \sigma )\, &\cir&
 \lambda_{\tau}(\tau ,\vec \sigma ),\qquad
 \partial_{\tau}\, \eta (\tau ,\vec \sigma )\, \cir\, A_{\tau}(\tau ,\vec \sigma
 ),\qquad \partial_{\tau}\, A_{\perp\, r}(\tau ,\vec \sigma )\, \cir\,
 - \pi_{\perp\, r}(\tau ,\vec \sigma ),\nonumber \\
 &&{}\nonumber \\
 &&dynamical\, Hamilton\, equations\nonumber \\
 &&{}\nonumber \\
 \partial_{\tau}\, \pi^r_{\perp}(\tau ,\vec \sigma )\, &\cir&\,
 \triangle\, A^r_{\perp}(\tau ,\vec \sigma ),\quad
 \Rightarrow\quad \Box\, A_{\perp\, r}(\tau ,\vec \sigma )\,
 \cir\, 0.
 \label{3.3}
 \eea
\bigskip

To fix the gauge we must only add a gauge fixing
$\varphi_{\eta}(\tau ,\vec \sigma ) \approx 0$ to the Gauss law,
which determines $\eta$. Its time constancy, i.e. $\partial_{\tau}\,
\varphi_{\eta}(\tau ,\vec \sigma ) + \{ \varphi_{\eta}(\tau ,\vec
\sigma ), H_D\} =  \varphi_{A_{\tau}}(\tau
,\vec \sigma ) \approx 0$, will generate the gauge fixing
$\varphi_{A_{\tau}}(\tau ,\vec \sigma ) \approx 0$ for $A_{\tau}$.
Finally the time constancy $\partial_{\tau}\,
\varphi_{A_{\tau}}(\tau ,\vec \sigma ) + \{ \varphi_{A_{\tau}}(\tau
,\vec \sigma ), H_D\} \approx 0$ will determine the Dirac multiplier
$\lambda_{\tau}(\tau ,\vec \sigma )$. By adding these {\it
two}  gauge fixing constraints  to the
first class constraints $\pi^{\tau}(\tau ,\vec \sigma ) \approx 0$,
$\Gamma (\tau ,\vec \sigma ) \approx 0$, one gets two pairs of
second class constraints allowing the elimination of the gauge
degrees of freedom so that only the DO's survive.

\subsection{Special Relativistic Killing Symmetries of the Electro-Magnetic Field}

Given a vector field

\beq
 X = \xi^A(\tau ,\vec \sigma )\,
\partial_A = \xi^{\tau}(\tau ,\vec \sigma )\,
\partial_{\tau} + \xi^r(\tau ,\vec \sigma )\, \partial_r,
 \label{3.4}
 \eeq

\noindent let us look for electro-magnetic potentials $A_A(\tau
,\vec \sigma )$ satisfying the Killing equations

\beq
 \varphi_A(\tau ,\vec \sigma ) = [L_X\, A]_A(\tau ,\vec \sigma ) =
 \xi^B(\tau ,\vec \sigma )\, \partial_B\, A_A(\tau ,\vec \sigma )
 + A_B(\tau ,\vec \sigma )\, \partial_A\, \xi^B(\tau ,\vec \sigma
 ) =0.
 \label{3.5}
 \eeq

In phase space the four equations $\varphi_A(\tau ,\vec \sigma ) =
0$ must be reinterpreted as {\it four constraints} added by hand and
restricting the configurations of the electro-magnetic field to
those having this Killing symmetry. Since the electro-magnetic gauge
theory has only two gauge degrees of freedom, the Killing equations
are also a restriction on the DO's ${\vec A}_{\perp}$,
${\vec \pi}_{\perp}$.

Moreover in phase space we have to ask (like for the  ordinary gauge
fixings) the time constancy of these (added by hand) four extra
constraints. Again a priori this will add other four conditions,
which have to be studied and again asked to be preserved in time
(and so on ..). Only at the end, if this procedure does not lead to
inconsistencies, we can say that the theory admits electro-magnetic
fields with the given Killing symmetry.

\bigskip

To rewrite the four Killing equations $\varphi_A(\tau ,\vec \sigma )
= 0$ as constraints we must use the first (kinematical) half of
Hamilton equations (\ref{3.3}) to replace the velocities
($\partial_{\tau}$ derivatives of the canonical variables) with
their phase space expression.

\medskip

Let us study in detail the four Killing equations (\ref{3.5}).\medskip

1) The $A = \tau$ Killing equation generates the following
constraint

\bea
 \varphi_{\tau}(\tau ,\vec \sigma ) &=& \Big[\xi^{\tau}\, \partial_{\tau}\, A_{\tau}
 + \xi^s\, \partial_s\, A_{\tau} + A_{\tau}\, \partial_{\tau}\, \xi^{\tau} +
 A_s\, \partial_{\tau}\, \xi^s\Big](\tau ,\vec \sigma )\,
 \cir\nonumber \\
 &\cir& \Big[\xi^{\tau}\, \lambda_{\tau} + \xi^s\, \partial_s\, A_{\tau} +
 A_{\tau}\, \partial_{\tau}\, \xi^{\tau} + (\partial_s\, \eta +
 A_{\perp\, s})\, \partial_{\tau}\, \xi^s\Big](\tau ,\vec \sigma )
 \approx 0,
 \label{3.6}
 \eea
\medskip

We have

\bea
 &&\{ \varphi_{\tau}(\tau ,\vec \sigma ), H_c\} =
 \partial_{\tau}\, \xi^s(\tau ,\vec \sigma )\, \{ A_{\perp\, s}(\tau ,\vec \sigma
 ), H_c\}\, \cir\, \partial_{\tau}\, \xi^s(\tau ,\vec \sigma )\,
 \pi_{\perp\, s}(\tau ,\vec \sigma ),\nonumber \\
 &&\{ \varphi_{\tau}(\tau ,\vec \sigma ), \pi^{\tau}(\tau ,{\vec \sigma}_1
 )\} = \xi^s(\tau ,\vec \sigma )\, \partial_s\, \delta^3(\vec
 \sigma - {\vec \sigma}_1) + \delta^3(\vec \sigma -{\vec
 \sigma}_1)\, \partial_{\tau}\, \xi^{\tau}(\tau ,\vec \sigma
 ),\nonumber \\
 &&\{ \varphi_{\tau}(\tau ,\vec \sigma ), \Gamma (\tau ,{\vec \sigma}_1
 )\} = - \partial_{\tau}\, \xi^s(\tau ,\vec \sigma )\,
 \partial_s\, \delta^3(\vec \sigma - {\vec \sigma}_1),\nonumber \\
 &&\{ \varphi_{\tau}(\tau ,\vec \sigma ), \int d^3\sigma_1\,
 [\lambda_{\tau}\, \pi^{\tau}](\tau ,{\vec \sigma}_1 )\} =
 \xi^s(\tau ,\vec \sigma )\, \partial_s\, \lambda_{\tau}(\tau ,\vec \sigma
 ),\nonumber \\
 &&\{ \varphi_{\tau}(\tau ,\vec \sigma ), - \int d^3\sigma_1\,
 [A_{\tau}\, \Gamma ](\tau ,{\vec \sigma}_1 ) = - \partial_{\tau}\,
 \xi^s(\tau ,\vec \sigma )\, \partial_s\, A_{\tau}(\tau ,\vec \sigma
 ),
 \label{3.7}
 \eea

 \medskip

The time constancy of $\varphi_{\tau}(\tau ,\vec \sigma ) \approx 0$
generates the extra constraint [the Killing constraint has an
explicit  $\tau$-dependence through the $\xi^A(\tau ,\vec \sigma )$
components of the Killing vector field and the Dirac multiplier
$\lambda_{\tau}(\tau ,\vec \sigma )$]

\bea
 \psi_{\tau}(\tau ,\vec \sigma ) &=& \partial_{\tau}\,
 \varphi_{\tau}(\tau ,\vec \sigma ) + \{ \varphi_{\tau}(\tau ,\vec \sigma
 ), H_D\} \cir\nonumber \\
 &\cir& \Big[\partial_{\tau}\, \xi^s\, \pi_{\perp\, s} + \xi^s\, \partial_s\,
 \lambda_{\tau} + \partial_{\tau}\, \xi^{\tau}\, \lambda_{\tau}
 + \xi^{\tau}\, \partial_{\tau}\, \lambda_{\tau} +\nonumber \\
 &+& A_{\tau}\, \partial^2_{\tau}\, \xi^{\tau} + (\partial_s\, \eta + A_{\perp\, s})
 \, \partial^2_{\tau}\, \xi^s\Big](\tau ,\vec \sigma ) \approx 0.
 \label{3.8}
 \eea

\bigskip

2) The $A = r$ Killing equations generate the following three
constraints

\bea
 \varphi_r(\tau ,\vec \sigma ) &=&  \Big[\xi^{\tau}\,
 \partial_{\tau}\, (\partial_r\, \eta + A_{\perp\, r}) +
 \xi^s\, \partial_s\, (\partial_r\, \eta + A_{\perp\, r})
 +\nonumber \\
 &+& A_{\tau}\, \partial_r\, \xi^{\tau} + (\partial_s\, \eta  +
 A_{\perp\, s})\, \partial_r\, \xi^s\Big](\tau ,\vec \sigma )\,
 \cir\nonumber \\
 &\cir& \Big[\xi^{\tau}\, (\partial_r\, A_{\tau} + \pi_{\perp\, r})
 + \partial_r\, (\xi^s\, \partial_s\, \eta ) + A_{\tau}\, \partial_r\,
 \xi^{\tau} +\nonumber \\
 &+& \xi^s\, \partial_s\, A_{\perp\, r} + A_{\perp\, s}\,
 \partial_r\, \xi^s\Big](\tau ,\vec \sigma ) \approx 0.
 \label{3.9}
 \eea

We have

\bea
 &&\{ \varphi_r(\tau ,\vec \sigma ), H_c\} = \Big[\xi^{\tau}\,
\triangle\, A_{\perp\, r} + \xi^s\, \partial_s\, \pi_{\perp\, r} +
\pi_{\perp\, s}\, \partial_r\, \xi^s\Big](\tau ,\vec \sigma ),
\nonumber \\
&&\{ \varphi_r(\tau ,\vec \sigma ), \int d^3\sigma_1\,
[\lambda_{\tau}\, \pi^{\tau}](\tau ,{\vec \sigma}_1 )\} =
\partial_r\, [\lambda_{\tau} \, \xi^{\tau}](\tau ,\vec \sigma ),
\nonumber \\
&&\{ \varphi_r(\tau ,\vec \sigma ), - \int d^3\sigma_1\, [A_{\tau}\,
\Gamma ](\tau ,{\vec \sigma}_1 )\} = - \partial_r\, [\xi^s\,
\partial_s\, A_{\tau}](\tau ,\vec \sigma ).
 \label{3.10}
\eea
\medskip

The time constancy of the Killing constraints $\varphi_r(\tau ,\vec
\sigma ) \approx 0$ generates the extra constraints (the dynamical
Hamilton equations (\ref{3.3}) are used)

\bea
 \psi_r(\tau ,\vec \sigma ) &=& \partial_{\tau}\, \varphi_r(\tau ,\vec \sigma )
 + \{ \varphi_r(\tau ,\vec \sigma ), H_D\} \cir\nonumber \\
 &\cir& \Big[\partial_{\tau}\, \xi^{\tau}\, (\partial_r\, A_{\tau} +
 \pi_{\perp\, r}) + \partial_r\, (\partial_{\tau}\, \xi^s\,
 \partial_s\, \eta ) + A_{\tau}\, \partial_r\, \partial_{\tau}\,
 \xi^{\tau} +\nonumber \\
 &+& \partial_{\tau}\, \xi^s\, \partial_s\, A_{\perp\, r} +
 A_{\perp\, s}\, \partial_r\, \partial_{\tau}\, \xi^s +
 \partial_r\, (\xi^{\tau}\, \lambda_{\tau} - \xi^s\, \partial_s\, A_{\tau})
+\nonumber \\
 &+& \xi^{\tau}\, \triangle\, A_{\perp\, r} + \xi^s\, \partial_s\,
 \pi_{\perp\, r} + \pi_{\perp\, s}\, \partial_r\, \xi^s
\Big](\tau ,\vec \sigma ) \approx 0.
 \label{3.11}
 \eea

\bigskip

Since we have already 8 constraints on the 6 existing variables, i.e. the 2 gauge variables
$A_{\tau}$, $\eta$, and the 4 DO's ${\vec a}_{\perp}$, ${\vec \pi}_{\perp}$, the constraints
$\psi_A(\tau, \vec \sigma) \approx 0$ must be identically conserved:
$\partial_{\tau}\, \psi_A(\tau ,\vec \sigma ) = 0$.

\subsection{A Time-like Killing Vector}

To understand the meaning of these Killing constraints let us
consider the  {\it time-translation Killing vector field} $X = \partial_{\tau}$
with $\xi^{\tau}(\tau ,\vec \sigma ) = 1$ and $\xi^s(\tau ,\vec
\sigma ) = 0$.

\medskip

In this case Eqs.(\ref{3.6}) and (\ref{3.3}) imply

\bea
 \varphi_{\tau}(\tau ,\vec \sigma ) &=& \lambda_{\tau}(\tau ,\vec \sigma )
 \, \cir\, \partial_{\tau}\, A_{\tau}(\tau ,\vec \sigma )\, \cir\,
 \partial^2_{\tau}\, \eta (\tau ,\vec \sigma )
 \approx 0,\qquad \psi_{\tau}(\tau ,\vec \sigma ) = \partial_{\tau}\,
 \lambda_{\tau}(\tau, \vec \sigma) \approx 0\, \nonumber \\
 &&{}\nonumber \\
  \varphi_r(\tau ,\vec \sigma ) &=& [\partial_r\, A_{\tau} +
 \pi_{\perp\, r}](\tau ,\vec \sigma )\, \cir\, [\partial_r\, A_{\tau} -
 \partial_{\tau}\, A_{\perp\, r}](\tau, \vec \sigma ) \approx 0,\nonumber  \\
 \psi_r(\tau ,\vec \sigma ) &=& \Big[\partial_r\, \lambda_{\tau} +
 \triangle\, A_{\perp\, r}\Big](\tau ,\vec \sigma ) \approx
 \triangle\, A_{\perp\, r}(\tau ,\vec \sigma ) \approx 0,\nonumber \\
 &&{}\nonumber \\
 \partial_r\, \varphi_r(\tau ,\vec \sigma ) &\approx& - \triangle\,
 A_{\tau}(\tau ,\vec \sigma ) \approx 0,\qquad \partial_r\,
\psi_r(\tau ,\vec \sigma )  = - \triangle\, \lambda_{\tau}(\tau
,\vec \sigma ) \approx 0,\nonumber \\
 \varphi_{\perp\, r}(\tau ,\vec \sigma ) &=& \pi_{\perp\, r}(\tau ,\vec \sigma )
 \approx 0,\qquad \psi_{\perp\, r}(\tau ,\vec \sigma ) = \triangle\,
 A_{\perp\, r}(\tau ,\vec \sigma ) \approx 0.
  \label{3.12}
 \eea

Therefore $\psi_{\tau}(\tau ,\vec\sigma ) \approx 0$ and
$\partial_r\, \psi_r(\tau ,\vec \sigma ) \approx 0$ do not
imply constraints being identically satisfied.
\medskip

The two Killing constraints $\varphi_{\tau}(\tau ,\vec \sigma )
\approx 0$ and $\partial_r\, \varphi_r(\tau ,\vec \sigma ) \approx
0$ are two gauge fixing constraints implying $\lambda_{\tau}(\tau
,\vec \sigma ) \approx 0$ and the following residual gauge freedom

\bea
 A_{\tau}(\tau ,\vec \sigma ) &\approx& A_{\tau}(\vec
 \sigma ),\qquad \triangle\, A_{\tau}(\vec \sigma ) \approx 0,\nonumber \\
 \eta (\tau ,\vec \sigma ) &\approx& \eta_o(\vec \sigma ) +
 \tau\, A_{\tau}(\vec \sigma ),
 \label{3.13}
 \eea

Therefore the gauge function $A_{\tau}(\vec \sigma )$ must be
harmonic. But, since the electro-magnetic potential is assumed to
vanish at spatial infinity, this means $A_{\tau}(\vec \sigma ) = 0$
and $\eta (\tau ,\vec \sigma ) \approx \eta_o(\vec \sigma )$. As a
consequence these two Killing constraints imply: i) the gauge fixing
constraint $A_{\tau}(\tau ,\vec \sigma ) \approx 0$ for
$\pi^{\tau}(\tau ,\vec \sigma ) \approx 0$; ii) the restriction of
the gauge fixing constraint for the Gauss law $\Gamma (\tau ,\vec
\sigma ) \approx 0$ to the form $\eta (\tau ,\vec \sigma ) -
\eta_o(\vec \sigma ) \approx 0$ with $\eta_o(\vec \sigma )$ an
arbitrary function. This is a family of gauges with only a residual
$\tau$-independent longitudinal gauge freedom and we get
$\varphi_r(\tau ,\vec \sigma ) = \pi_{\perp\, r}(\tau ,\vec \sigma )
 \, \cir\, - \partial_{\tau}\, A_{\perp\, r}(\tau ,\vec \sigma )
 \approx 0$, namely $A_{\perp\, r}(\tau ,\vec \sigma ) = A_{\perp\,
 r}(\vec \sigma )$.
 \medskip

The remaining four Killing constraints correspond to the transverse
parts of $\varphi_r(\tau ,\vec \sigma ) \approx 0$ and $\psi_r(\tau
,\vec \sigma ) \approx 0$, i.e. they are $\varphi_{\perp\, r}(\tau
,\vec \sigma ) = \pi_{\perp\, r}(\tau ,\vec \sigma ) \approx 0$ and
$\psi_{\perp\, r}(\tau ,\vec \sigma ) = \triangle\, A_{\perp\,
r}(\vec \sigma ) \approx 0$. Again the boundary conditions at
spatial infinity imply that the harmonic functions  $A_{\perp\,
r}(\vec \sigma )$ vanish. Therefore these Killing constraints form
the two pairs of second class constraints $\pi_{\perp\, r}(\tau
,\vec \sigma ) \approx 0$, $A_{\perp\, r}(\tau ,\vec \sigma )
\approx 0$ killing the DO's of the electro-magnetic
field, i.e. its transverse radiative components. Finally the consistency
conditions $\partial_{\tau}\, \psi_A(\tau,\vec \sigma ) \approx 0$ are identically satisfied.

\bigskip

In conclusion, with the Killing vector field $X = \partial_{\tau}$
the allowed set of electro-magnetic configurations is composed only
by {\it pure gauge configurations}. This leaves space only for the introduction of static magnetic
monopoles. The addition of another Killing vector field will only reduce the
residual gauge freedom of this pure longitudinal gauge
configuration.

\bigskip

One can study the effect of the imposition of the other nine Killing symmetries on the electro-magnetic
field in the same way. While a Killing symmetry associated with a spatial translation in direction
"i" can be shown to imply that the electro-magnetic field does not depend on $\sigma^i$, a rotational
Killing symmetry implies electro-magnetic fields rotationally invariant around an axis. Finally it can
be shown that the Killing symmetry under a boost in direction "i" implies that there is only a static
magnetic field produced by a potential $A_{\perp a}(\sigma^{b \not= i})$ and no electric field: therefore
there is no genuine radiation field.

\section{Tetrad Gravity in Asymptotically Minkowskian Space-Times}

After a review of canonical ADM tetrad gravity \cite{16,17,18,19,20}
we will study the Hamiltonian constraints implied by a Killing symmetry.

\subsection{The York Canonical Basis for ADM Tetrad Gravity}

In ADM tetrad gravity \cite{16,17,18,19,20}
the 4-scalar 4-metric defined in Section II is decomposed on cotetrads ${}^4g_{AB}(\tau, \vec \sigma) = E_A^{(\alpha)}(\tau, \vec \sigma)\,
{}^4\eta_{(\alpha)(\beta)}\, E^{(\beta)}_B(\tau, \vec \sigma)$; they are the dynamical variables.
The associated tetrads ${}^4E^A_{(\alpha)}(\tau, \vec
\sigma)$  ($(\alpha )$ are flat indices) are connected with the world tetrads
${}^4E^{\mu}_{(\alpha)}(\tau, \vec \sigma) = z^{\mu}_A(\tau, \vec
\sigma)\, {}^4E^A_{(\alpha)}(\tau, \vec \sigma)$ by using
the embedding $z^{\mu}(\tau ,\vec \sigma)$ of the
instantaneous 3-spaces. As said in Section II we have $z^{\mu}_{\tau} = (1 + n)\, l^{\mu} + N^r\, z^{\mu}_r$
with $N^r = n_{(a)}\, {}^3e^r_{(a)}$, where ${}^re^r_{(a)}$ are triads on the 3-space $\Sigma_{\tau}$.
The tetrads admit the following decomposition

\beq
 {}^4E^A_{(\alpha )} = {}^4{\buildrel \circ \over {\bar E}}^A_{(o)}\,
 L^{(o)}{}_{(\alpha )}(\varphi_{(c)}) + {}^4{\buildrel \circ \over
 {\bar E}}^A_{(b)}\, R^T_{(b)(a)}(\alpha_{(c)})\,
 L^{(a)}{}_{(\alpha )}(\varphi_{(c)}),
 \label{4.1}
 \eeq

\noindent where $\varphi_{(a)}(\tau ,\vec \sigma )$ and $\alpha_{(a)}(\tau ,\vec \sigma )$
are the boost and rotation variables of the O(3,1) gauge freedom in the choice of the tetrads
and of their transport. The following barred variables are independent from the angles $\alpha_{(a)}$:
${\bar n}_{(a)} = \sum_b\, n_{(b)}\, R_{(b)(a)}(\alpha_{(e)})$,
${}^3e^r_{(a)} = R_{(a)(b)}(\alpha_{(e)})\, {}^3{\bar e}^r_{(b)}$.

\medskip

In Eqs.(\ref{4.1}) there are the following tetrads and cotetrads
adapted to the chosen 3-space $\Sigma_{\tau}$ ($l_A = z^{\mu}_A\, l_{\mu}$)

\bea
 {}^4{\buildrel \circ \over {\bar E}}^A_{(o)}
 &=&  {1\over {1 + n}}\, (1; - {\bar n}_{(a)}\,
 {}^3{\bar e}^r_{(a)}) = l^A,\qquad {}^4{\buildrel \circ \over
 {\bar E}}^A_{(a)} = (0; {}^3{\bar e}^r_{(a)}), \nonumber \\
 &&{}\nonumber  \\
 {}^4{\buildrel \circ \over {\bar E}}^{(o)}_A
 &=&  (1 + n)\, (1; \vec 0) = \sgn\, l_A,\qquad
 {}^4{\buildrel \circ \over {\bar E}}^{(a)}_A
= ({\bar n}_{(a)}; {}^3{\bar e}_{(a)r}).
 \label{4.2}
 \eea

\medskip

As shown in Refs. \cite{16,19} the natural configuration variables of ADM tetrad gravity are $\varphi_{(a)}$, $n$, $n_{(a)}$,
${}^3e_{(a)r}$. The conjugate momenta are $\pi_{\varphi_{(a)}}$, $\pi_n$, $\pi_{n_{(a)}}$, ${}^3\pi^r_{(a)}$. There are 14 (ten primary and four secondary)
first-class constraints: seven of the primary ones are $\pi_{\varphi_{(a)}} \approx 0$, $\pi_n \approx 0$, $\pi_{n_{(a)}} \approx 0$.
\medskip

In Ref.\cite{19}  a canonical basis adapted to all the 10 primary first-class constraints was found
with a Shanmugadhasan canonical transformation. It leads to the following York canonical basis
(see Ref.\cite{19} for the boundary conditions at spatial infinity of the canonical variables)

\bea
 &&\begin{minipage}[t]{3cm}
\begin{tabular}{|l|l|l|l|} \hline
$\varphi_{(a)}$ & $n$ & $n_{(a)}$ & ${}^3e_{(a)r}$ \\ \hline
$\pi_{\varphi_{(a)}} \approx 0$ & $\pi_n \approx 0$ & $
\pi_{n_{(a)}} \approx 0 $ & ${}^3{ \pi}^r_{(a)}$
\\ \hline
\end{tabular}
\end{minipage} \hspace{1cm}\nonumber \\
 &&{}\nonumber \\
 &&{\longrightarrow \hspace{.2cm}} \
\begin{minipage}[t]{4 cm}
\begin{tabular}{|ll|ll|l|l|l|} \hline
$\varphi_{(a)}$ & $\alpha_{(a)}$ & $n$ & ${\bar n}_{(a)}$ &
$\theta^r$ & $\tilde \phi$ & $R_{\bar a}$\\ \hline
$\pi_{\varphi_{(a)}} \approx0$ &
 $\pi_{\alpha_{(a)}} \approx 0$ & $\pi_n \approx 0$ & $\pi_{{\bar n}_{(a)}} \approx 0$
& $\pi^{(\theta )}_r$ & $\pi_{\tilde \phi}$ & $\Pi_{\bar a}$ \\
\hline
\end{tabular}
\end{minipage}\nonumber \\
 &&{}
 \label{4.3}
 \eea

The secondary first-class constraints are the super-Hamiltonian and super-momentum ones: they are
partial differential equations for the determination of $\tilde \phi$ and $\pi^{(\theta)}_r$ in
terms of $\theta^r$, $\pi_{\tilde \phi}$, $R_{\bar a}$, $\Pi_{\bar a}$.
\medskip

Due to the use of radar 4-coordinates all
the canonical variables of the York basis are 4-scalars of the space-time,
but they have different 3-tensorial behaviors inside the 3-spaces. $\theta^i$ and $\pi_{\tilde \phi}$
are the primary inertial gauge variables, while $n$ and ${\bar n}_{(a)}$ are the secondary ones.
\bigskip

In the York canonical basis we have (from now on we will use
$V_{ra}$ for $V_{ra}(\theta^n)$ to simplify the notation)\footnote{The set of numerical parameters $\gamma_{\bar aa}$
satisfies $\sum_u\, \gamma_{\bar au} = 0$, $\sum_u\,
\gamma_{\bar a u}\, \gamma_{\bar b u} = \delta_{\bar a\bar b}$,
$\sum_{\bar a}\, \gamma_{\bar au}\, \gamma_{\bar av} = \delta_{uv} -
{1\over 3}$. Each solution of these equations defines a different
York canonical basis.}

\begin{eqnarray*}
  Q_a\, &=& e^{\sum_{\bar a}^{1,2}\, \gamma_{\bar aa}\, R_{\bar a}}, \qquad
 \tilde \phi = \sqrt{det\, {}^3g},\qquad
 \pi_{\tilde \phi} =   {{c^3}\over {12\pi\, G}}\, {}^3K,\nonumber \\
 {}&&\nonumber \\
 {}^3e_{(a)r} &=& \sum_b\, R_{(a)(b)}(\alpha_{(e)})\, {}^3{\bar
 e}_{(b)r},\qquad {}^3{\bar e}_{(a)r} = {\tilde \phi}^{1/3}\, Q_a\,
 V_{ra},\nonumber \\
 {}^3e^r_{(a)} &=& \sum_b\, R_{(a)(b)}(\alpha_{(e)})\, {}^3{\bar
 e}^r_{(b)},\qquad {}^3{\bar e}^r_{(a)} = {\tilde \phi}^{- 1/3}\, Q^{-1}_a\,
 V_{ra},
 \end{eqnarray*}

\bea
 {}^4g_{\tau\tau} &=& \sgn\, \Big[(1 + n)^2 - \sum_a\,
 {\bar n}_{(a)}^2\Big],\nonumber \\
 {}^4g_{\tau r} &=& - \sgn\, \sum_a\, {\bar n}_{(a)}\, {}^3{\bar e}_{(a)r} =
 - \sgn\, {\tilde \phi}^{1/3}\, \sum_a\, Q_a\,
 V_{ra}\, {\bar n}_{(a)},\nonumber \\
 {}^4g_{rs} &=& - \sgn\, {}^3g_{rs}
 =  - \sgn\, \sum_a\, {}^3{\bar e}_{(a)r}\, {}^3{\bar e}_{(a)s} =
  - \sgn\, {\tilde \phi}^{2/3}\,
 \sum_a\, Q^2_a\, V_{ra}\, V_{sa},\nonumber \\
 &&{}^3g^{rs} = {\tilde \phi}^{-2/3}\,
 \sum_a\, Q^{-2}_a\, V_{ra}\, V_{sa},\nonumber \\
 {}^4g^{\tau\tau} &=& {{\sgn}\over {(1 + n)^2}},\qquad
  {}^4g^{\tau r} = -\sgn\, {\tilde \phi}^{-1/3}\,  {{Q_a^{-1}\, V_{ra}\,
  {\bar n}_{(a)}}\over {(1 + n)^2}},\nonumber \\
 {}^4g^{rs} &=&  -\sgn\, {\tilde \phi}^{-2/3}\, Q_a^{-1}\,
 Q_b^{-1}\, V_{ra}\, V_{sb}\, (\delta_{(a)(b)} -
 {{{\bar n}_{(a)}\, {\bar n}_{(b)}}\over {(1 + n)^2}}).
 \label{4.4}
 \eea

\bigskip

$\alpha_{(a)}(\tau ,\vec \sigma )$ and $\varphi_{(a)}(\tau ,\vec
\sigma )$ are the 6 configuration variables parametrizing the O(3,1)
gauge freedom in the choice of the tetrads in the tangent plane to
each point of $\Sigma_{\tau}$ and describe the arbitrariness in the
choice of a tetrad to be associated to a time-like observer, whose
world-line goes through the point $(\tau ,\vec \sigma )$. They fix
{\it the unit 4-velocity of the observer and the conventions for the
orientation of gyroscopes and their transport along the world-line
of the observer}. The gauge variables $\theta^i(\tau, \vec \sigma)$, $n(\tau, \vec
\sigma)$, ${\bar n}_{(a)}(\tau, \vec \sigma)$ describe inertial
effects, which are the the relativistic counterpart of the
non-relativistic ones (the centrifugal, Coriolis,... forces in
Newton mechanics in accelerated frames) and which are present also
in the non-inertial frames of Minkowski space-time \cite{25}.

\medskip

In Eq.(\ref{4.4}) the quantity ${}^3K(\tau, \vec \sigma)$ is the
trace of the extrinsic curvature ${}^3K_{rs}(\tau, \vec \sigma)$ of
the instantaneous 3-spaces $\Sigma_{\tau}$ whose expression is given in Appendix A.
This so-called York time ${}^3K(\tau ,\vec \sigma)$ is the only gauge variable among the momenta: this
is a reflex of the Lorentz signature of space-time, because
$\pi_{\tilde \phi}(\tau ,\vec \sigma)$ and $\theta^n(\tau ,\vec \sigma)$ can be used as a set of
4-coordinates \cite{19,20,31}. Its conjugate variable, to be determined by the super-Hamiltonian
constraint, is the conformal factor of the 3-metric $\tilde \phi(\tau ,\vec \sigma)$, which
is the {\it 3-volume density} on $\Sigma_{\tau}$: $V_R
= \int_R d^3\sigma\, \tilde \phi(\tau ,\vec \sigma)$, $R \subset \Sigma_{\tau}$.

\medskip

The two pairs of canonical variables $R_{\bar a}(\tau ,\vec \sigma)$, $\Pi_{\bar a}(\tau ,\vec \sigma)$,
$\bar a = 1,2$, describe the generalized {\it tidal effects}, namely
the independent degrees of freedom of the gravitational field. In
particular the configuration tidal variables $R_{\bar a}$ depend
{\it only on the eigenvalues of the 3-metric}. They are DO
 {\it only} with respect to the gauge transformations
generated by 10 of the 14 first class constraints. Let us remark
that, if we fix completely the gauge and we go to Dirac brackets,
then the only surviving dynamical variables $R_{\bar a}$ and
$\Pi_{\bar a}$ become two pairs of {\it non canonical} DO for that gauge
(see Ref.\cite{21}  for new results on the DO's). \medskip

The Dirac Hamiltonian is (the $\lambda$'s are arbitrary Dirac
multipliers)

\bea
 H_D &=& E_{ADM} + \int d^3\sigma\, \Big[- \sgn\, c\, n\, {\cal H} + {\bar n}_{(a)}\,
 {\bar {\cal H}}_{(a)}\Big](\tau ,\vec \sigma ) +\nonumber \\
 &+& \int d^3\sigma\, \Big[\lambda_n\, \pi_n + \lambda_{{\bar n}_{(a)}}\,
 \pi_{{\bar n}_{(a)}} + \lambda_{\varphi_{(a)}}\, \pi_{\varphi_{(a)}}
 + \lambda_{\alpha_{(a)}}\, \pi_{\alpha_{(a)}} \Big](\tau ,\vec \sigma ).
 \label{4.5}
 \eea

\noindent See Eqs. (3.45), (B.8), (3.42) and (3.44) of the first
paper in Ref.\cite{20} for the expression of the ADM energy $E_{ADM}$ and of the super-Hamiltonian,
$\cal H$, and super-momentum, ${\bar {\cal H}}_{(a)}$, constraints: all these quantities are functions of
$\theta^r$, $\pi_r^{(\theta)}$, $\tilde \phi$, $\pi_{\tilde \phi}$, $R_{\bar a}$, $\Pi_{\bar a}$, but not of $n$ and ${\bar n}_{(a)}$.

\bigskip

In the following we shall work in the {\it Schwinger time gauge}
$\varphi_{(a)}(\tau, \vec \sigma) \approx 0$ (tetrads adapted to the
3+1 splitting) and $\alpha_{(a)}(\tau,  \vec \sigma) \approx 0$
(arbitrary choice of an origin for rotations), where ${}^3e_{(a)r}(\tau ,\vec \sigma) \approx
{}^3{\bar e}_{(a)r}(\tau ,\vec \sigma)$, $\lambda_{\varphi_{(a)}}(\tau ,\vec \sigma) =
\lambda_{\alpha_{(a)}}(\tau ,\vec \sigma) = 0$.

\bigskip

The first kinematical half of Hamilton equations implies the following expressions for the $\partial_{\tau}$
derivatives (the velocities)

\bea
  \partial_{\tau}\, n(\tau ,\vec \sigma ) &\cir&  \lambda_n(\tau ,\vec \sigma ),\nonumber \\
 \partial_{\tau}\, {\bar n}_{(a)}(\tau ,\vec \sigma ) &\cir& \lambda_{ {\bar n}_{(a)}}(\tau ,\vec \sigma ),\nonumber \\
 \partial_{\tau}\, {}^3{\bar e}_{(a)r}(\tau ,\vec \sigma ) &\cir& \Big[-
 (1 + n)\, {}^3K_{rs}\, {}^3{\bar e}^s_{(a)} + \partial_r\, {\bar
 n}_{(a)} + {\bar n}_{(b)}\, {}^3{\bar e}^s_{(b)}\, (\partial_s\, {}^3{\bar e}_{(a)r} -
 \partial_r\, {}^3{\bar e}_{(a)s}) \Big](\tau ,\vec \sigma ),\nonumber \\
 &&{}\nonumber \\
 \partial_{\tau}\, {}^4g_{\tau\tau}(\tau ,\vec \sigma ) &\cir& 2\,
 \sgn\, \Big[(1 + n)\, \lambda_n - {\bar n}_{(a)}\,
  \lambda_{ {\bar n}_{(a)}}\Big](\tau ,\vec \sigma ),\nonumber \\
 \partial_{\tau}\, {}^4g_{\tau r}(\tau ,\vec \sigma ) &\cir& -
 \sgn\, \Big[\lambda_{ {\bar n}_{(a)}}\, {}^3{\bar e}_{(a)r} + {\bar n}_{(a)}\,
 \Big(- (1 + n)\, {}^3K_{rs}\, {}^3{\bar e}^s_{(a)} + \nonumber \\
 &+& \partial_r\, {\bar n}_{(a)} + {\bar n}_{(b)}\, {}^3{\bar e}^s_{(b)}\, (\partial_s\,
 {}^3{\bar e}_{(a)r} - \partial_r\, {}^3{\bar e}_{(a)s})
 \Big)\Big](\tau ,\vec \sigma ),\nonumber \\
 \partial_{\tau}\, {}^4g_{rs}(\tau ,\vec \sigma ) &\cir&
 \Big[\partial_r\, ({\bar n}_{(a)}\, {}^3{\bar e}_{(a)s}) +
 \partial_s\, ({\bar n}_{(a)}\, {}^3{\bar e}_{(a)r}) -
 2\, {}^3\Gamma^u_{rs}\, {\bar n}_{(a)}\, {}^3{\bar e}_{(a)u} -
 2\, (1 + n)\, {}^3K_{rs}\Big](\tau ,\vec \sigma ).\nonumber \\
 &&{}
 \label{4.6}
 \eea

\subsection{The Killing Equations Associated to the Given Killing
Vector Field $X$.}

We shall assume that the 4-metric is left
invariant by a given Killing vector field $X = \xi^A(\tau ,\vec
\sigma )\, \partial_A$: $L_X\, {}^4g_{AB}(\tau ,\vec \sigma )\, d\sigma^A\,
d\sigma^B = 0$. With generic tetrads one has  $L_X\, {}^4E^{(\alpha )}_A(\tau ,\vec \sigma
)\, d\sigma^A \not= 0$. However as shown in Ref. \cite{32} and in its bibliography, the
existence of the Killing vector $X$ for the 4-metric implies that there is a special set of tetrads
${}^4{\tilde E}^{(\alpha )}_A(\tau ,\vec \sigma )$
such that $L_X\, {}^4{\tilde E}^{(\alpha )}_A(\tau ,\vec
\sigma )\, d\sigma^A\, =\, 0$.

\bigskip

The existence of the Killing vector implies the 10 Killing equations

\beq
 \chi_{AB}(\tau ,\vec \sigma ) =
 \Big({}^4\nabla_A\, \xi_B + {}^4\nabla_B\, \xi_A\Big)(\tau ,\vec \sigma ) = \Big(\partial_A\, \xi_B
+ \partial_B\, \xi_A - 2\, {}^4\Gamma^C_{AB}\, \xi_C\Big)(\tau ,\vec \sigma ) = 0.
 \label{4.7}
 \eeq

\bigskip

By using the notation of the previous Subsection  we have

\bea
 &&\xi_A = {}^4g_{AB}\, \xi^B,\qquad \xi^A = {}^4g^{AB}\,
 \xi_A,\nonumber \\
 &&{}\nonumber \\
 \xi_{\tau} &=& \sgn\, \Big[\Big((1 + n)^2 - {\bar n}_{(a)}\,
 {\bar n}_{(a)}\Big)\, \xi^{\tau} - {\bar n}_{(a)}\, {}^3{\bar e}_{(a)r}\,
 \xi^r\Big],\nonumber \\
 \xi_r &=& -\sgn\, \Big[{\bar n}_{(a)}\, {}^3{\bar e}_{(a)r}\, \xi^{\tau} +
 {}^3g_{rs}\, \xi^s\Big],\nonumber \\
 &&{}\nonumber \\
 \xi^{\tau} &=& {{\sgn}\over {(1 + n)^2}}\, \Big[\xi_{\tau} -
 {\bar n}_{(a)}\, {}^3{\bar e}^r_{(a)}\, \xi_r\Big],\nonumber \\
 \xi^r &=& -\sgn\, {}^3{\bar e}^r_{(a)}\, \Big[{}^3{\bar e}^s_{(a)}\, \xi_s +
 {{{\bar n}_{(a)}\, (\xi_{\tau} - {\bar n}_{(b)}\, {}^3{\bar e}^s_{(b)}\, \xi_s)}\over
 {(1 + n)^2}}\Big].
 \label{4.8}
 \eea

\medskip

By using Eqs.(\ref{4.6}) for the time-derivative of the metric and Eq.(\ref{4.4}) for its spatial derivatives
($\partial_r\, {}^4g_{\tau\tau} \cir 2 \sgn\, \Big[(1 + n)\,
\partial_r\, n - {\bar n}_{(a)}\, \partial_r\, {\bar n}_{(a)}\Big]$,
$\partial_s\, {}^4g_{\tau r} = -\sgn\, \partial_s\, ({\bar n}_{(a)}\,
{}^3{\bar e}_{(a)r})$,  $\partial_u\, {}^4g_{rs} = -\sgn\, \partial_u\, {}^3g_{rs} =
-\sgn\, ({}^3{\bar e}_{(a)r}\, \partial_u\, {}^3{\bar e}_{(a)s} +
{}^3{\bar e}_{(a)s}\, \partial_u\, {}^3{\bar e}_{(a)r})$ ), we get

 \bea
 \partial_{\tau}\, \xi_{\tau} &\cir& \sgn\, \Big[ \Big((1 +
 n)^2 - {\bar n}_{(a)}\, {\bar n}_{(a)}\Big)\, \partial_{\tau}\, \xi^{\tau} -
 {\bar n}_{(a)}\, {}^3{\bar e}_{(a)r}\, \partial_{\tau}\, \xi^r +\nonumber \\
 &+& 2\, \Big((1 + n)\, \lambda_n - {\bar n}_{(a)}\,
\lambda_{ {\bar n}_{(a)}} \Big)\, \xi^{\tau} -
  \Big(\lambda_{ {\bar n}_{(a)}}\, {}^3e_{(a)r} +
  {\bar n}_{(a)}\, \Big[ \partial_r\, {\bar n}_{(a)} +\nonumber \\
 &+& {\bar n}_{(b)}\, {}^3{\bar e}^s_{(b)}\, (\partial_s\, {}^3{\bar e}_{(a)r} -
 \partial_r\, {}^3{\bar e}_{(a)s}) - (1 + n)\, {}^3K_{rv}\,
 {}^3{\bar e}^v_{(a)}\Big]\Big)\, \xi^r\Big],\nonumber \\
 &&{}\nonumber \\
 &&{}\nonumber \\
 \partial_r\, \xi_{\tau} &=& \sgn\, \Big[\Big((1 + n)^2 -
 {\bar n}_{(a)}\, {\bar n}_{(a)}\Big)\, \partial_r\, \xi^{\tau} - {\bar n}_{(a)}\,
 {}^3{\bar e}_{(a)s}\, \partial_r\, \xi^s +\nonumber \\
 &+& 2\, \Big((1 + n)\, \partial_r\, n - {\bar n}_{(a)}\,
 \partial_r\, {\bar n}_{(a)}\Big)\, \xi^{\tau} - \partial_r\, ({\bar n}_{(a)}\,
 {}^3{\bar e}_{(a)s})\, \xi^s\Big],
 \label{4.9}
 \eea

 \bea
 \partial_{\tau}\, \xi_r &\cir&  -\sgn\, \Big[{\bar n}_{(a)}\,
 {}^3{\bar e}_{(a)r}\, \partial_{\tau}\, \xi^{\tau} + {}^3g_{rs}\,
 \partial_{\tau}\, \xi^s + \nonumber \\
 &+& \Big(\lambda_{ {\bar n}_{(a)}}\, {}^3{\bar e}_{(a)r}
 + {\bar n}_{(a)}\, \Big[ \partial_r\, {\bar n}_{(a)} + {\bar n}_{(b)}\,
 {}^3{\bar e}^s_{(b)}\, (\partial_s\, {}^3{\bar e}_{(a)r} - \partial_r\,
 {}^3{\bar e}_{(a)s}) - (1 + n)\, {}^3K_{rv}\, {}^3{\bar e}^v_{(a)}\Big]\Big)\,
 \xi^{\tau} +\nonumber \\
 &+& \Big({\bar n}_{(a)}\, \Big[\partial_r\, {}^3{\bar e}_{(a)s} + \partial_s\,
 {}^3{\bar e}_{(a)r} - 2\, {}^3\Gamma^u_{rs}\, {}^3{\bar e}_{(a)u}\Big] +
  \partial_r\, {\bar n}_{(a)}\, {}^3{\bar e}_{(a)s} + \partial_s\, {\bar n}_{(a)}
 \, {}^3{\bar e}_{(a)r} - 2\, (1 + n)\, {}^3K_{rs} \Big)\, \xi^s\Big],
 \nonumber \\
 &&{}\nonumber \\
 &&{}\nonumber \\
 \partial_r\, \xi_s &=& -\sgn\, \Big[ {\bar n}_{(a)}\, {}^3{\bar e}_{(a)s}\,
 \partial_r\, \xi^{\tau} + {}^3g_{sv}\, \partial_r\, \xi^v
 + \partial_r\, ({\bar n}_{(a)}\, {}^3{\bar e}_{(a)s})\, \xi^{\tau} +
 ({}^3{\bar e}_{(a)s}\, \partial_r\, {}^3{\bar e}_{(a)v} + {}^3{\bar e}_{(a)v}\,
 \partial_r\, {}^3{\bar e}_{(a)s})\, \xi^v\Big].\nonumber \\
 &&{}
 \label{4.10}
 \eea

\subsection{The Hamiltonian Expression of the Killing Equations.}

By using Eqs.(\ref{a1}) and (\ref{4.8}) we get the following
expression for the 10 Killing constraints implied by the Killing equations (\ref{4.7})

\bea
 {1\over 2}\, \chi_{\tau\tau} &=& \partial_{\tau}\, \xi_{\tau} -
 ({}^4\Gamma^{\tau}_{\tau\tau}\, \xi_{\tau} +
 {}^4\Gamma^u_{\tau\tau}\, \xi_u) \cir\nonumber \\
 &&{}\nonumber \\
 &\cir& \sgn\, \Big[ \Big((1 + n)^2 - {\bar n}_{(a)}\, {\bar n}_{(a)}\Big)\,
 \partial_{\tau}\, \xi^{\tau} - {\bar n}_{(a)}\, {}^3{\bar
 e}_{(a)r}\, \partial_{\tau}\, \xi^r +\nonumber \\
 &+& \Big[(1 + n)\, \lambda_n - {\bar n}_{(a)}\,
 \lambda_{ {\bar n}_{(a)}}\Big]\, \xi^{\tau}
 +\nonumber \\
 &+& \Big((1 + n)\, {}^3{\bar e}_{(a)r}\, \Big[{}^3{\bar e}^s_{(a)}\, \partial_s\, n -
 {}^3{\bar K}_{(a)(b)}\, {\bar n}_{(b)}\Big] -\nonumber \\
 &-& {\bar n}_{(a)}\, \Big[\partial_r\, {\bar n}_{(a)} + {\bar n}_{(b)}\,
 {}^3{\bar e}^s_{(b)}\,(\partial_s\, {}^3{\bar e}_{(a)r} - \partial_s\, {}^3{\bar e}_{(a)s})\Big]\,
 \Big)\, \xi^r \Big] \approx 0,
 \label{4.11}
 \eea

\bea
 \chi_{\tau r} &=& \partial_{\tau}\, \xi_r + \partial_r\,
 \xi_{\tau} - 2\, ({}^4\Gamma^{\tau}_{\tau r}\, \xi_{\tau} +
 {}^4\Gamma^u_{\tau r}\, \xi_u) \cir\nonumber \\
 &&{}\nonumber \\
 &\cir& \sgn\, \Big[\Big((1 + n)^2 - {\bar n}_{(a)}\, {\bar n}_{(a)}\Big)\,
 \partial_r\, \xi^{\tau} - {\bar n}_{(a)}\, {}^3{\bar e}_{(a)s}\,
 \partial_r\, \xi^s -\nonumber \\
 &-& {\bar n}_{(a)}\, {}^3{\bar e}_{(a)r}\, \partial_{\tau}\,
 \xi^{\tau} - {}^3g_{rs}\, \partial_{\tau}\, \xi^s -\nonumber \\
 &-& \Big({}^3{\bar e}_{(a)r}\, \lambda_{ {\bar n}_{(a)}} + {\bar n}_{(a)}\,
 \partial_r\, {\bar n}_{(a)} + (1 + n)\, {}^3{\bar K}_{r(a)}\, {\bar n}_{(a)}
 +\nonumber \\
 &+& {\bar n}_{(a)}\, {\bar n}_{(b)}\, {}^3{\bar e}^s_{(b)}\, (\partial_s\,
 {}^3{\bar e}_{(a)r} - \partial_r\, {}^3{\bar e}_{(a)s})\Big)\, \xi^{\tau} -
  \partial_s\, \Big({\bar n}_{(a)}\, {}^3{\bar e}_{(a)r}\Big)\, \xi^s
 \Big]\approx 0,\nonumber \\
 &&{}
 \label{4.12}
 \eea

\bea
 \chi_{rs} &=& \partial_r\, \xi_s + \partial_s\, \xi_r - 2\,
 ({}^4\Gamma^{\tau}_{rs}\, \xi_{\tau} + {}^4\Gamma^u_{rs}\, \xi_u)
 \cir\nonumber \\
 &&{}\nonumber \\
 &\cir& - \sgn\, \Big[{\bar n}_{(a)}\, {}^3{\bar e}_{(a)s}\,
 \partial_r\, \xi^{\tau} + {}^3g_{sv}\, \partial_r\, \xi^v +
 {\bar n}_{(a)}\, {}^3{\bar e}_{(a)r}\,
 \partial_s\, \xi^{\tau} + {}^3g_{rv}\, \partial_s\, \xi^v
 +\nonumber \\
 &+& \Big[- 2\, (1 + n)\, {}^3K_{rs} +
 \partial_r\, ({\bar n}_{(a)}\, {}^3{\bar e}_{(a)s}) +
\partial_s\, ({\bar n}_{(a)}\, {}^3{\bar e}_{(a)r}) -
\nonumber \\
&-& {\bar n}_{(a)}\, {}^3{\bar e}^v_{(a)}\, (\partial_r\, {}^3g_{sv}
+ \partial_s\, {}^3g_{rv} - \partial_v\, {}^3g_{rs})
 \Big]\, \xi^{\tau} - \xi^v\, \partial_v\, {}^3g_{rs} \Big] \approx 0,
 \label{4.13}
 \eea

\bigskip

By consistency we must have

\beq
 \psi_{AB}(\tau ,\vec \sigma ) = \partial_{\tau}\, \chi_{AB}(\tau ,\vec \sigma ) +
 \{ \chi_{AB}(\tau ,\vec \sigma ), H_D\},
\approx 0,
 \label{4.14}
 \eeq

\noindent where we have to use Eqs.(\ref{4.5}) for the Dirac Hamiltonian with
$\lambda_{\varphi_{(a)}}(\tau ,\vec \sigma ) = \lambda_{\alpha_{(a)}}(\tau ,\vec \sigma ) = 0$
(Schwinger time gauges) and where $\partial_{\tau}$ acts on $\xi^A(\tau ,\vec \sigma )$.

to evaluate the velocities.

\bigskip

In the Schwinger time gauges the 16 variables consisting in
the 8 gauge variables $n$, ${\bar n}_{(a)}$, $\theta^r$, $\pi_{\tilde \phi}$,
in the 4 physical tidal variables $R_{\bar a}$, $\Pi_{\bar a}$, and
in the 4 Dirac multipliers $\lambda_n$, $\lambda_{{\bar n}_{(a)}}$
(with $\tilde \phi$ and $\pi_r^{(\theta)}$  determined by the secondary first-class constraints
${\cal H}(\tau ,\vec \sigma ) \approx 0$, ${\bar {\cal H}}_{(a)}(\tau ,\vec \sigma )
\approx 0$ as functions of $\theta^r$, $\pi_{\tilde \phi}$, $R_{\bar a}$, $\Pi_{\bar a}$)
are restricted by the 20 Killing constraints $\chi_{AB}(\tau ,\vec \sigma ) \approx 0$, $\psi_{AB}(\tau ,\vec \sigma ) \approx 0$.
Therefore some of  these Killing constraints must be void not to have over-determined equations
and moreover this implies that we must have $\partial_{\tau}\, \psi_{AB}(\tau ,\vec \sigma) = 0$ automatically satisfied.

\subsection{A Time-like Killing Vector}

Let us try to understand the meaning of these Killing constraints for $X = \partial_{\tau}$
({\it stationary} space-times). We have $\xi^{\tau} = 1$ and $\xi^r
= 0$, so that $\xi_{\tau} = \sgn\, \Big((1 + n)^2 - {\bar n}_{(a)}\,
{\bar n}_{(a)}\Big)$, $\xi_r = -\sgn\, {\bar n}_{(a)}\, {}^3{\bar e}_{(a)r}$.

\bea
 \chi_{\tau\tau} &\cir& 2\, \sgn\, \Big( (1 + n)\, \lambda_n
 - n_{(a)}\, \lambda_{ {\bar n}_{(a)}}\Big) \approx 0,
 \label{4.15}
 \eea

 \bea
 \chi_{\tau r} &\cir& - \sgn\, \Big( {}^3{\bar e}_{(a)r}\,
 \lambda_{ {\bar n}_{(a)}} + {\bar n}_{(a)}\,
 \partial_r\, {\bar n}_{(a)} +\nonumber \\
 &+& {\bar n}_{(a)}\, {\bar n}_{(b)}\, {}^3{\bar e}^v_{(b)}\,
 (\partial_v\, {}^3{\bar e}_{(a)r} - \partial_r\, {}^3{\bar e}_{(a)v})
 - (1 + n)\, {}^3K_{rv}\, {}^3{\bar e}^v_{(a)}\Big) \approx 0,
 \label{4.16}
 \eea

 \bea
 \chi_{rs} &=& -\sgn\, \Big(- 2\, (1 + n)\, {}^3K_{rs} + \partial_r\, ({\bar n}_{(a)}\,
 {}^3{\bar e}_{(a)s}) + \partial_s\, ({\bar n}_{(a)}\, {}^3{\bar e}_{(a)r}) -\nonumber \\
 &-& {\bar n}_{(a)}\, {}^3{\bar e}^v_{(a)}\, (\partial_r\, {}^3g_{sv} +
 \partial_s\, {}^3g_{rv} - \partial_v\, {}^3g_{rs})\Big) \approx 0.
 \label{4.17}
 \eea

As a consequence we get (${}^3{\bar K}_{(a)(b)} = {}^3{\bar e}^r_{(a)}\, {}^3{\bar
e}^s_{(b)}\, {}^3K_{rs}$, ${}^3{\bar K}_{r(a)} = {}^3{\bar
e}^s_{(a)}\, {}^3K_{rs}$)

\bea
 \lambda_n &\cir& \partial_{\tau}\, n \approx {{{\bar n}_{(a)}\, 
 \lambda_{ {\bar n}_{(a)}} }\over {1 + n}},\nonumber \\
 \lambda_{ {\bar n}_{(a)}} &\cir& \partial_{\tau}\, {\bar
 n}_{(a)} \approx - {}^3{\bar e}^r_{(a)}\, \Big[{\bar n}_{(b)}\, \partial_r\, {\bar n}_{(b)}
 + (1 + n)\, {}^3{\bar K}_{r(b)}\, {\bar n}_{(b)} +
{\bar n}_{(b)}\, {\bar n}_{(c)}\, {}^3{\bar e}^v_{(c)}\,
 (\partial_v\, {}^3{\bar e}_{(b)r} - \partial_r\, {}^3{\bar e}_{(b)v})
 \Big],\nonumber \\
 {}^3K_{rs} &\approx& {1\over {2\, (1 + n)}}\, \Big[\partial_r\, ({\bar n}_{(a)}\,
 {}^3{\bar e}_{(a)s}) + \partial_s\, ({\bar n}_{(a)}\, {}^3{\bar e}_{(a)r})
 -\nonumber \\
 &-& {\bar n}_{(a)}\, {}^3{\bar e}^v_{(a)}\, (\partial_r\, {}^3g_{sv} +
 \partial_s\, {}^3g_{rv} - \partial_v\, {}^3g_{rs})\Big],
 \label{4.18}
 \eea

\noindent with ${}^3g_{rs} = {\tilde \phi}^2/3\, \sum_a\, Q^2_a\, V_{ra}(\theta^u)\, V_{sa}(\theta^u)$ from Eq.(\ref{4.4})
and with ${}^3K_{rs}$ given in Eq.(\ref{a1}) as a function of the canonical variables in the York canonical basis.

\bigskip

As in the electro-magnetic case, the Killing constraints $\chi_{\tau
A} \approx 0$ restrict the gauge freedom by determining the 4 Dirac
multipliers $\lambda_n$ and $\lambda_{ {\bar n}_{(a)}}$
associated to lapse and shift.

\medskip

The extra constraints $\psi_{\tau A}
\approx 0$ will be automatically satisfied (like in the
electro-magnetic case) since they determine the velocities
$\partial_{\tau}\, \lambda_n$ and $\partial_{\tau}\, \lambda_{ {\bar n}_{(a)}}$
of the already determined Dirac multipliers.

\medskip

The traces of the constraints $\chi_{rs} \approx 0$ and $\psi_{rs} \approx 0$
determine the gauge variable ${}^3K$ ($\pi_{\tilde \phi}$), namely
the clock synchronization convention, and the lapse function $n$, namely they determine
the primary and secondary gauge variables associated with the Dirac multiplier $\lambda_n$.
Like in the electro-magnetic case some residual $\tau$-independent gauge freedom can be left
by the chosen boundary conditions at spatial infinity for the tetrads.

\medskip

The 10 constraints from the traceless part of the constraints $\chi_{rs} \approx 0$ and $\psi_{rs} \approx 0$
are the equations for the determination of
the 3 primary $\theta^r$ and 3 secondary ${\bar n}_{(a)}$ gauge variables
associated with the Dirac multiplier $\lambda_{{\bar n}_{(a)}}$ and
of the four DO's   $R_{\bar a}$, $\Pi_{\bar a}$.

\medskip

Therefore no physical tidal variables (no gravitational waves in the linearized theory) survive to
the presence of a time-like Killing symmetry. Only a $\tau$-independent gauge freedom is left and only static singularities like black holes
are allowed.

With non-time-like Killing symmetries  tidal variables adapted to the symmetry  would survive.

\section{Conclusions}

In this paper the lacking Hamiltonian formulation of Killing symmetries
in terms of Dirac constraints added by hand was given. This was done both for
the electro-magnetic field and for tetrad gravity in asymptotically Minkowskian
space-times.

It was shown that in both cases the presence of a time-like Killing symmetry kills all the
physical degrees of freedom, namely the DO's. This result was known to many people but
is not present in the literature as far as we know.

\appendix

\section{Hamiltonian Expressions}

The Hamiltonian expression \cite{20,21} of the extrinsic curvature ${}^3K_{rs}$ of the instantaneous 3-spaces $\Sigma_{\tau}$
and of the 4-Christoffel symbols ${}^4\Gamma^A_{BC} = {1\over 2}\, {}^4g^{AE}\, \Big(\partial_B\,
 {}^4g_{CE} + \partial_C\, {}^4g_{BE} - \partial_E\, {}^4g_{BC}\Big)$ is

\begin{eqnarray*}
 {}^3{\tilde K}_{rs} &=& \sgn\, {{4\pi\, G}\over {c^3}}\, {\tilde \phi}^{-1/3}\, \Big(\sum_a\,
Q^2_a\, V_{ra}(\theta^n)\, V_{sa}(\theta^n)\, [2\, \sum_{\bar b}\,
\gamma_{\bar ba}\, \Pi_{\bar b} -  \tilde \phi\,
\pi_{\tilde \phi}] +\nonumber \\
 &+& \sum_{ab}\, Q_a\, Q_b\, [V_{ra}(\theta^n)\, V_{sb}(\theta^n) +
 V_{rb}(\theta^n)\, V_{sa}(\theta^n)]\, \sum_{twi}\, {{\epsilon_{abt}\,
 V_{tw}(\theta^n)\, B_{iw}(\theta^n)\, \pi_i^{(\theta )}}\over {
 Q_b\, Q^{-1}_a  - Q_a\, Q^{-1}_b}} \Big),
 \end{eqnarray*}

\begin{eqnarray*}
 {}^4\Gamma^{\tau}_{\tau\tau} &=& {1\over {1 + n}}\,
 \Big(\partial_{\tau}\, n + {\bar n}_{(a)}\, {}^3{\bar e}^r_{(a)}\, \partial_r\,
 n - {\bar n}_{(a)}\, {\bar n}_{(b)}\, {}^3{\bar K}_{(a)(b)}\Big) \cir \nonumber \\
 &\cir& {1\over {1 + n}}\,
 \Big(\lambda_n + {\bar n}_{(a)}\, {}^3{\bar e}^r_{(a)}\, \partial_r\,
 n - {\bar n}_{(a)}\, {\bar n}_{(b)}\, {}^3{\bar K}_{(a)(b)}\Big)
 ,\nonumber \\
 &&{}\nonumber \\
 {}^4\Gamma^{\tau}_{\tau r} &=& {1\over {1 + n}}\, \Big(\partial_r\, n -
 {}^3{\bar K}_{r(a)}\, {\bar n}_{(a)}\Big),\nonumber \\
 &&{}\nonumber \\
 {}^4\Gamma^{\tau}_{rs} &=& - {1\over {1 + n}}\, {}^3K_{rs},\nonumber \\
 &&{}\nonumber \\
 {}^4\Gamma^u_{\tau\tau} &\cir& {}^3{\bar e}^u_{(a)}\,
 \Big[\partial_{\tau}\, {\bar n}_{(a)} - {{{\bar n}_{(a)}}\over {1 + n}}\,
 \partial_{\tau}\, n +\nonumber \\
 &+& (1 + n)\, \Big(\delta_{(a)(b)} - {{{\bar n}_{(a)}\, {\bar n}_{(b)}}\over
 {(1 + n)^2}}\Big)\, \Big({}^3{\bar e}^r_{(b)}\, \partial_r\, n -
 {}^3{\bar K}_{(b)(c)}\, {\bar n}_{(c)}\Big)\Big] \cir\nonumber \\
 &\cir& {}^3{\bar e}^u_{(a)}\,
 \Big[{\bar \lambda}_{ {\vec {\bar n}}(a)} - {{{\bar n}_{(a)}}\over {1 + n}}\,
 \lambda_n +\nonumber \\
 &+& (1 + n)\, \Big(\delta_{(a)(b)} - {{{\bar n}_{(a)}\, {\bar n}_{(b)}}\over
 {(1 + n)^2}}\Big)\, \Big({}^3{\bar e}^r_{(b)}\, \partial_r\, n -
 {}^3{\bar K}_{(b)(c)}\, {\bar n}_{(c)}\Big)\Big],
 \end{eqnarray*}
 
 \bea
 {}^4\Gamma^u_{\tau r} &=& {}^3{\bar e}^u_{(a)}\,
 \Big[\partial_r\, {\bar n}_{(a)} - {{{\bar n}_{(a)}}\over {1 + n}}\,
 \partial_r\, n + {}^3{\bar \omega}_{r(a)(b)}\, {\bar n}_{(b)} -\nonumber \\
 &-& (1 + n)\, \Big(\delta_{(a)(b)} - {{{\bar n}_{(a)}\, {\bar n}_{(b)}}\over
 {(1 + n)^2}}\Big)\, {}^3{\bar K}_{r(b)}\Big] =\nonumber \\
 &=& {}^3{\bar e}^u_{(a)}\,
 \Big[\partial_r\, {\bar n}_{(a)} - {{{\bar n}_{(a)}}\over {1 + n}}\,
 \partial_r\, n - (1 + n)\, \Big(\delta_{(a)(b)} - {{{\bar n}_{(a)}\, {\bar n}_{(b)}}\over
 {(1 + n)^2}}\Big)\, {}^3{\bar K}_{r(b)}\Big] +\nonumber \\
 &+& {\bar n}_{(a)}\, \Big(\partial_r\, {}^3{\bar e}^u_{(a)} +
 {}^3\Gamma^u_{rs}\, {}^3{\bar e}^s_{(a)}\Big),\nonumber \\
 &&{}\nonumber \\
 {}^4\Gamma^u_{rs} &=& {}^3\Gamma^u_{rs} + {{{\bar n}_{(a)}}\over {1 + n}}\,
 {}^3{\bar e}^u_{(a)}\, {}^3K_{rs},\nonumber \\
 \qquad && {}^3\Gamma^u_{rs} = {1\over 2}\, {}^3g^{uv}\,
 \Big(\partial_r\, {}^3g_{sv} + \partial_s\,
 {}^3g_{rv} - \partial_v\, {}^3g_{rs}\Big).
 \label{a1}
 \eea

\vfill\eject

\end{document}